\documentclass[aps,prl,reprint,superscriptaddress]{revtex4-2}

\usepackage{amsmath}
\usepackage{graphicx} 
\usepackage{gensymb}
\usepackage{siunitx}
\usepackage{soul}
\usepackage{color}

\graphicspath{{figures/}}
\begin{document}

%\preprint{}

\title{Forced Imbibition in Stratified Porous Media}



\author{Nancy B. Lu}
\thanks{These authors contributed equally.}
\affiliation{Department of Chemical and Biological Engineering, Princeton University, Princeton, NJ 08544}
\author{Amir A. Pahlavan }
\thanks{These authors contributed equally.}
\affiliation{Department of Mechanical and Aerospace Engineering, Princeton University, Princeton, NJ 08544}
\author{Christopher A. Browne}
\affiliation{Department of Chemical and Biological Engineering, Princeton University, Princeton, NJ 08544}
\author{Daniel B. Amchin}
\affiliation{Department of Chemical and Biological Engineering, Princeton University, Princeton, NJ 08544}
\author{Howard A. Stone}
\affiliation{Department of Mechanical and Aerospace Engineering, Princeton University, Princeton, NJ 08544}
\author{Sujit S. Datta}
\email[To whom correspondence should be addressed: \\]{ssdatta@princeton.edu}
%\thanks{To whom correspondence should be addressed:\\ssdatta@princeton.edu}
\affiliation{Department of Chemical and Biological Engineering, Princeton University, Princeton, NJ 08544}


\date{\today}

\begin{abstract}
Imbibition plays a central role in diverse energy, environmental, and industrial processes. In many cases, the medium has multiple parallel strata of different permeabilities; however, how this stratification impacts imbibition is poorly understood. We address this gap in knowledge by directly visualizing forced imbibition in three-dimensional (3D) porous media with two parallel strata. We find that imbibition is spatially heterogeneous: for small capillary number Ca, the wetting fluid preferentially invades the fine stratum, while for Ca above a threshold value, the fluid instead preferentially invades the coarse stratum. This threshold value depends on the medium geometry, the fluid properties, and the presence of residual wetting films in the pore space. These findings are well described by a linear stability analysis that incorporates crossflow between the strata. Thus, our work provides quantitative guidelines for predicting and controlling flow in stratified porous media.
\end{abstract}

%\keywords{}

\maketitle

%%%%%%%%%%%%%%% INTRODUCTION
\section{Introduction}
Imbibition, the process by which a wetting fluid displaces a non-wetting fluid from a porous medium, plays crucial roles in our lives. It underlies key energy processes, such as oil/gas recovery \cite{morrow2001recovery, mattax1962imbibition, li2000characterization,nicolaides2015impact} and water management in fuel cells \cite{forner2016advanced}; environmental processes, such as geological CO$_2$ sequestration \cite{bennion2006supercritical,bennion2010drainage, hatiboglu2008pore,celia2015status},  groundwater aquifer remediation \cite{schaefer2000experimental,esposito2011remediation}, and moisture infiltration in soil and wood \cite{weisbrod2002imbibition,tesoro2007relative,hassanein2006investigation}; and diverse other applications including operation of chemical reactors \cite{attou1999two} and wicking in fabrics \cite{tang2015characterizing,tang2015characterizing2}, paper microfluidics 
\cite{castro2017characterizing}, building materials \cite{hall,rahman2015recycled,takahashi1996acoustic}, and diagnostic devices \cite{hong2015dynamics}. As a result, the physics of imbibition has been studied widely. For homogeneous media with disordered pores of a single mean size, it is now understood how different invasion dynamics and flow patterns can arise depending on the pore size, solid wettability, surface roughness, as well as the fluid viscosity, interfacial tension, and flow boundary conditions \cite{lenormand1984role, chang2009experimental, hatiboglu2008pore, sun2016micro, hughes2000pore, lenormand1988numerical, lenormand1990liquids, sahimi1993flow, alava2004imbibition, stokes1986interfacial, weitz1987dynamic, hultmark2011influence,zhao2016wettability,tanino2018oil,odier2017forced}.

However, in many cases, porous media are not homogeneous. Instead, they can have parallel strata, characterized by different mean pore sizes, oriented along the direction of macroscopic fluid flow \cite{bear2013dynamics,galloway2012terrigenous,gasda2005upscaling,king2018microstructural}; these can further impact the interfacial dynamics in complex ways \cite{datta2013drainage,king2018microstructural,reyssat2009imbibition,zhou1997scaling,zapata1981theoretical,bear2013dynamics, galloway2012terrigenous,gasda2005upscaling}. Being able to predict and control imbibition in stratified porous media is therefore both fundamentally interesting and of critical practical importance. Unfortunately, the multi-scale nature of this problem poses a challenge: it involves a complex interplay between pore-scale capillary forces driving imbibition in the different strata, transverse viscous forces due to crossflow between strata, and longitudinal viscous forces due to macroscopic flow through the medium. A scaling analysis of these forces suggests that heterogeneous invasion behaviors can arise at different values of the capillary number Ca $\equiv\mu_{w}(Q/A)/\gamma$ \cite{zhou1997scaling}, which quantifies the relative importance of viscous and capillary forces at the pore scale; $\mu_{w}$ is the wetting fluid dynamic shear viscosity, $Q$ is the volumetric flow rate, $A$ is the overall cross-sectional area of the medium, and $\gamma$ is the interfacial tension between the fluids. However, systematic study of these dynamics is lacking, and thus the physics of imbibition in stratified porous media remains poorly understood. Some studies have indicated that heterogeneous invasion can arise during spontaneous imbibition from a bulk fluid reservoir \cite{ashraf2017spontaneous,ashraf2019capillary}, but did not investigate the distinct case of forced imbibition at a fixed flow rate, and overlooked the possible influence of residual wetting films---both key characteristics of many applications \cite{mcphee2015core}. Other studies have provided tantalizing evidence that different heterogeneous invasion behaviors can arise during forced imbibition, but only probed a narrow range of flow rates \cite{cinar2004experimental, dong2005immiscible, dawe1992experimental}, focused on bulk fluid saturation \cite{cinar2004experimental,dong1998characterization,zapata1981theoretical}, explored small differences in pore size between strata \cite{dawe1992experimental,cinar2004experimental, ahmed1988experimental,zapata1981theoretical,yokoyama1981effects,debbabi2017viscous,Debbabi2017, zapata1981theoretical}, or did not provide detailed characterization of residual wetting films \cite{zapata1981theoretical,dong2005immiscible,cinar2004experimental}.  Hence, a full understanding of forced imbibition in stratified porous media remains lacking. 

Here, we use confocal microscopy to directly visualize forced imbibition in stratified 3D porous media. We find dramatically different invasion behaviors depending on the imposed Ca. For sufficiently small Ca, the wetting fluid preferentially invades the fine stratum; by contrast, above a threshold Ca, the wetting fluid preferentially invades the coarse stratum instead. The threshold value depends on the medium geometry, the fluid properties, and the presence of residual wetting films in the pore space. Our experimental results can be rationalized by a linear stability analysis that considers the balance of capillary and viscous forces, along with crossflow between the strata, during the incipient stage of invasion. Together, these results elucidate the heterogeneous dynamics of imbibition, and provide quantitative guidelines for predicting and controlling flow in stratified porous media.

\begin{figure}[htp!]
\centering
    \includegraphics[width=\linewidth]{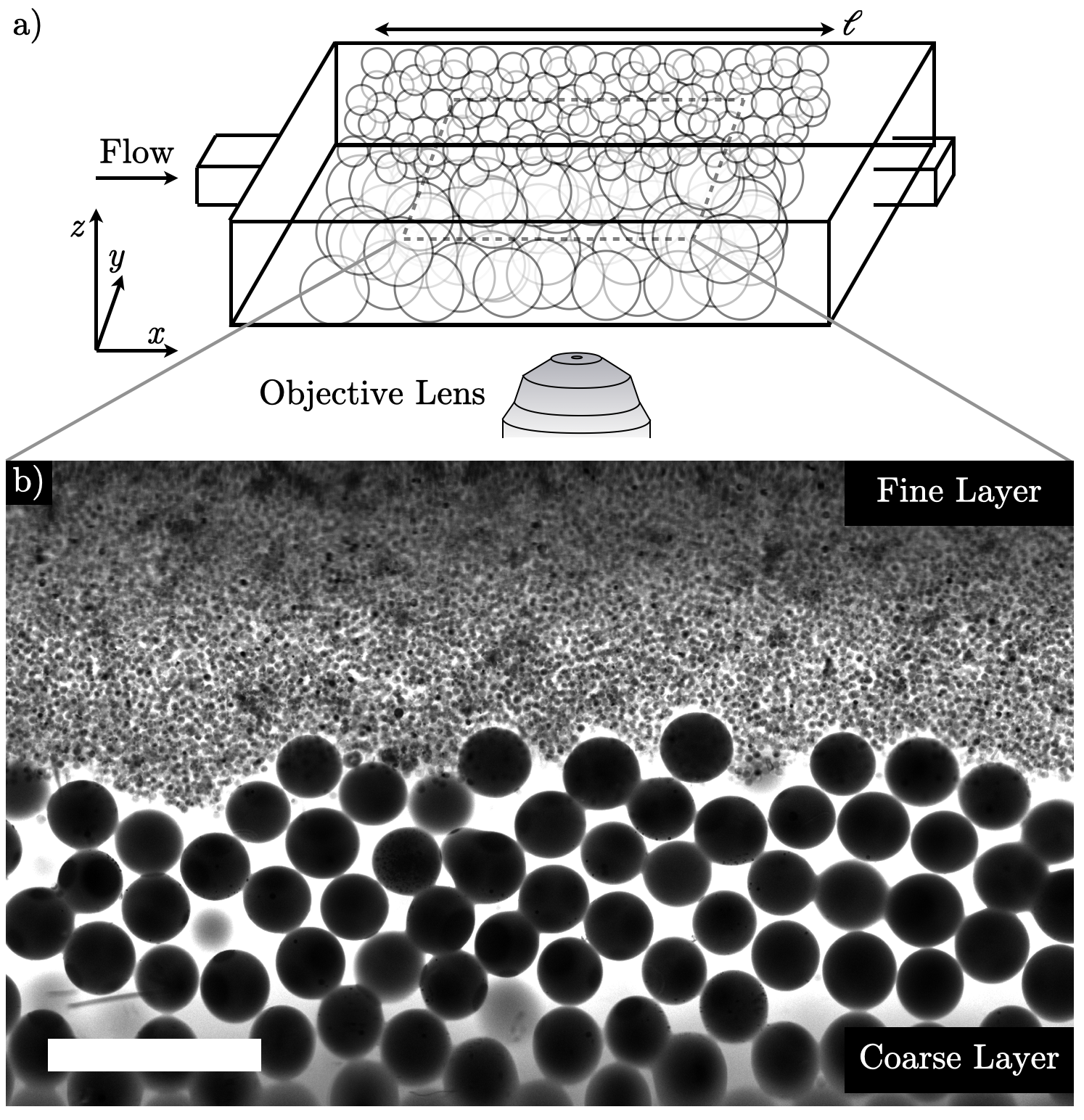}
    \caption{Overview of experiments in stratified porous media. (a) Schematic of the experimental setup. We use confocal microscopy to visualize imbibition in 3D stratified media composed of sintered packings of glass beads of different sizes. (b) Optical section acquired within a stratified medium. The pore space is saturated with the fluorescently-dyed wetting fluid, and the black circles show cross-sections of the beads making up the medium. Scale bar denotes 1 mm. }
    \label{fig:setup}\end{figure}

%%%%%%%%%%%%%%% EXPERIMENTAL METHODS
\section{Experimental Method}
We construct each porous medium by lightly sintering densely-packed borosilicate glass beads in a thin-walled square quartz capillary with total cross sectional area $A=$ 4 or 9 mm$^2$, following our previous work \cite{krummel2013visualizing,datta2013drainage}. As shown in Figure \ref{fig:setup}, we arrange beads of different mean diameters in parallel coarse and fine strata having different properties denoted by the subscripts $c$ and $f$ respectively, with the interface between the strata running along the direction of imposed flow. The mean bead diameters are $d_{c}=330\pm30~\mu$m and $d_{f}=42\pm4~\mu$m, and a slight polydispersity in the bead sizes results in the formation of a disordered packing in each stratum. The strata have the same length $\ell$ varying from 1.4 to 6.5 cm, porosity $\phi\approx41\%$ \cite{krummel2013visualizing}, and cross-sectional areas $A_{c}$ and $A_{f}$ with $A_c/A_f$ varying between 0.2 to 4. The difference in the bead sizes results in a large difference in the absolute permeabilities of the strata; we estimate these using the Kozeny-Carman equation previously validated for bead packings, $k_i=\phi_i^3d_i^2/180(1-\phi_i)^2$ \cite{philipse93,krummel2013visualizing}, where $i\in\{c,f\}$ denotes the stratum, yielding $k_{c}\approx120~\mu$m$^2$ and $k_{f}=1.9~\mu$m$^{2}$. The permeability ratio is then $k_{c}/k_{f}\approx60$, comparable to the permeability ratio characterizing many naturally-occurring strata \cite{king2018microstructural}.

To enable imaging of imbibition within the media, we use fluids whose compositions are carefully chosen to match their refractive indices with that of the glass beads, minimizing light scattering from the solid surfaces. The wetting fluid is a mixture of 91.6 wt\% dimethyl sulfoxide (DMSO) and 8.4 wt\% water with $\mu_w$ = 2.7 mPa-s. We dye the wetting fluid with 0.005 wt\% Rhodamine 6G to visualize it using laser fluorescence. The non-wetting fluid is a mixture of aromatic and aliphatic hydrocarbons (\textit{Cargille} refractive index liquids) with dynamic shear viscosity $\mu_{nw}$ = 16.8 mPa-s, also formulated to match its refractive index with that of the wetting fluid and the glass beads. The interfacial tension between the wetting and non-wetting fluids is $\gamma\approx 13$ mN/m, and the three-phase contact angle between the wetting fluid and glass in the presence of the non-wetting fluid is $\theta\approx5^\circ$ \cite{krummel2013visualizing}; we therefore take $\cos\theta\approx1$.

During each experiment, we displace undyed non-wetting fluid from the pore space by injecting the dyed wetting fluid at a constant flow rate $Q$, corresponding to a fixed capillary number Ca. To visualize the dynamics of the fluid displacement, we use confocal fluorescence microscopy to continually acquire successive images, each with a lateral area of $3169~\mu$m $\times$ $3169~\mu$m at a depth $\sim300~\mu$m within the medium, which span the entire cross-section of the medium.  The imaging depth corresponds to $\sim10$ close-packed beads deep within the fine stratum, and we assume that our measurements of interfacial dynamics are not influenced by the boundaries, which likely impact the flow at the walls of the capillary; assessing the influence of any possible boundary effects would be an important direction for future work. The brightness in the images   (e.g., Fig. \ref{fig:setup}b) reflects the fluorescent signal from the dyed wetting fluid as it invades the pore space; the undyed non-wetting fluid appears dark, and we identify the glass beads by their contrast with the dyed wetting fluid, as shown by the micrograph in Fig. \ref{fig:setup}b.

\section{Primary Imbibition}

%%%%%%%%%%%%%%% PRIMARY IMBIBITION NO CROSSFLOW
\begin{figure*}[htp!]
    \centering
    \includegraphics[width=\linewidth]{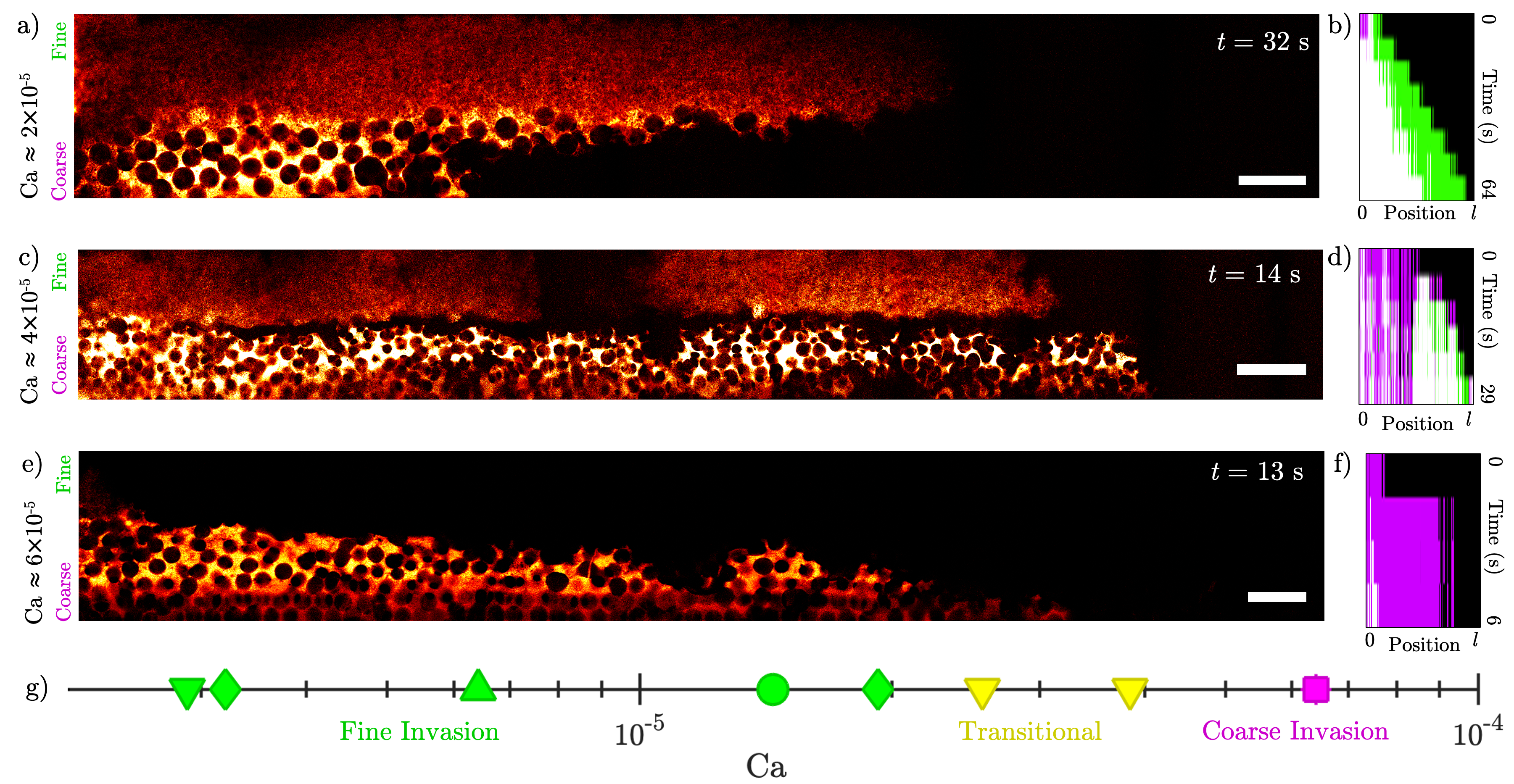}
    \caption{Heterogeneous invasion behavior during primary imbibition. Panels (a), (c), and (e) show optical sections through the entirety of two stratified porous media, obtained a time $t$ after the wetting fluid begins to invade each medium. The red region shows the dyed wetting fluid while the black shows the undyed non-wetting fluid, which initially fully saturates the pore space. Imposed flow direction is from left to right. Scale bars denote 1 mm. Panels (b), (d), and (f) are kymographs showing the positions at which the fine and coarse strata are invaded in green and magenta, respectively, while positions at which both strata are invaded are shown in white, as they vary over time. At a low capillary number Ca, wetting fluid preferentially invades the fine stratum, as shown in (a-b). At a moderate Ca, the wetting fluid simultaneously invades both strata, as shown in (c-d). At a sufficiently large Ca, the wetting fluid instead preferentially invades the coarse stratum, as shown in (e-f). (e) Summary of primary imbibition experiments performed at different Ca; green and magenta points indicate preferential invasion into the fine and coarse strata, respectively, while yellow points indicates that both strata are invaded at similar speeds. Specifically, to identify the yellow points, we measure the initial speed of the fluid-fluid interface in each stratum and apply the quantitative criterion that an experiment is in the transitional regime when the speed difference is $\leq20\%$ the speed of the faster moving interface. A magenta square and a magenta cross are overlapping. Circles, squares, diamonds, down-pointing triangles, up-pointing triangles, and crosses indicate $0.2\leq A_c/A_f < 0.7$ and $1.4\leq \ell < 2.0$ cm, $0.2\leq A_c/A_f < 0.7$ and $2.0\leq \ell < 3.0$ cm, $0.7\leq A_c/A_f < 1.5 $ and $1.4\leq \ell < 2.0$ cm, $0.7\leq A_c/A_f < 1.5 $ and $2.0\leq \ell < 3.0$ cm, $0.7\leq A_c/A_f < 1.5 $ and $ \ell \geq 3.0$ cm, and $ A_c/A_f \geq 1.5$ and $2.0\leq \ell < 3.0$ cm, respectively.}
    \label{primary_65x}
\end{figure*}
% a) Time point 37 at 194.26 seconds if when the wetting fluid hits the medium. Time point 38 at 199.56 seconds is the first snapshot in the figure. Time point 48 at 253.25 seconds is when the wetting fluid reaches the end of the fine stratum, so the velocity is 0.316 mm/s. Time point 50 at 263.91 seconds is when the fluid reaches the end of the coarse stratum, so the velocity is 0.2676 mm/s. 
% b) The wetting enters the medium between time point 16 (98.79 seconds) and 17 (105.30 seconds). Enters coarse stratum at time point 17 (105.30 seconds) and leaves at time point 18 (111.78 seconds), so the velocity is 2.541 mm/s. Enters fine stratum at time point 19 (118.17 seconds) and leaves at time point 27 (170.33) but very unclear, so the possible velocity is 0.474 mm/s. The length of the medium is 24.735 mm

We first investigate the case of primary imbibition at Ca $\approx2\times10^{-5}$, for which the medium is initially fully saturated with the non-wetting fluid, and therefore does not contain residual wetting films prior to imbibition. We investigate the influence of such films in the next section. The invasion dynamics are strongly heterogeneous: the wetting fluid enters both strata, but it preferentially invades the fine stratum, as shown by the micrograph in Fig. \ref{primary_65x}a. We quantify these dynamics by tracking the region invaded in each stratum over time, shown by the kymograph in Fig. \ref{primary_65x}b; the positions at which the fine and coarse strata are invaded are shown in green and magenta, respectively, while positions at which both strata are invaded are shown in white. As indicated by the straight boundaries in the kymograph, the wetting fluid invades each stratum at nearly constant speed, with the fine stratum invaded approximately twice as fast.

To further explore the dynamics of imbibition, we repeat this experiment at a larger Ca. For a transition region $2\times10^{-5}\lesssim\text{Ca}\lesssim6\times10^{-5}$, shown by the yellow points, both strata are invaded at similar speeds, as shown in Fig. \ref{primary_65x}c-d. Further, at even larger Ca $\approx6 \times 10^{-5}$, we observe strikingly different invasion dynamics: in this case, the wetting fluid preferentially invades the coarse stratum, as shown in Fig. \ref{primary_65x}e. In stark contrast to imbibition at Ca $=2 \times 10^{-5}$, here the wetting fluid invades the coarse stratum $\sim5$  times faster than it does the fine stratum, as shown by the kymograph in Fig. \ref{primary_65x}f. Thus, the heterogeneous invasion behavior underlying imbibition appears to dramatically change around a threshold $\text{Ca}$ between 2 and $6 \times 10^{-5}$. We confirm this expectation using further experiments performed at varying Ca, $A_c/A_f$, and $\ell$. For small $\text{Ca}<2\times10^{-5}$, the wetting fluid preferentially invades the fine stratum, as shown by the green points in Fig. \ref{primary_65x}g; conversely, for sufficiently large $\text{Ca}>6\times10^{-5}$, the wetting fluid preferentially invades the coarse stratum, as shown by the magenta points. 

\begin{figure}[htp!]
    \centering
    \includegraphics[width=\linewidth]{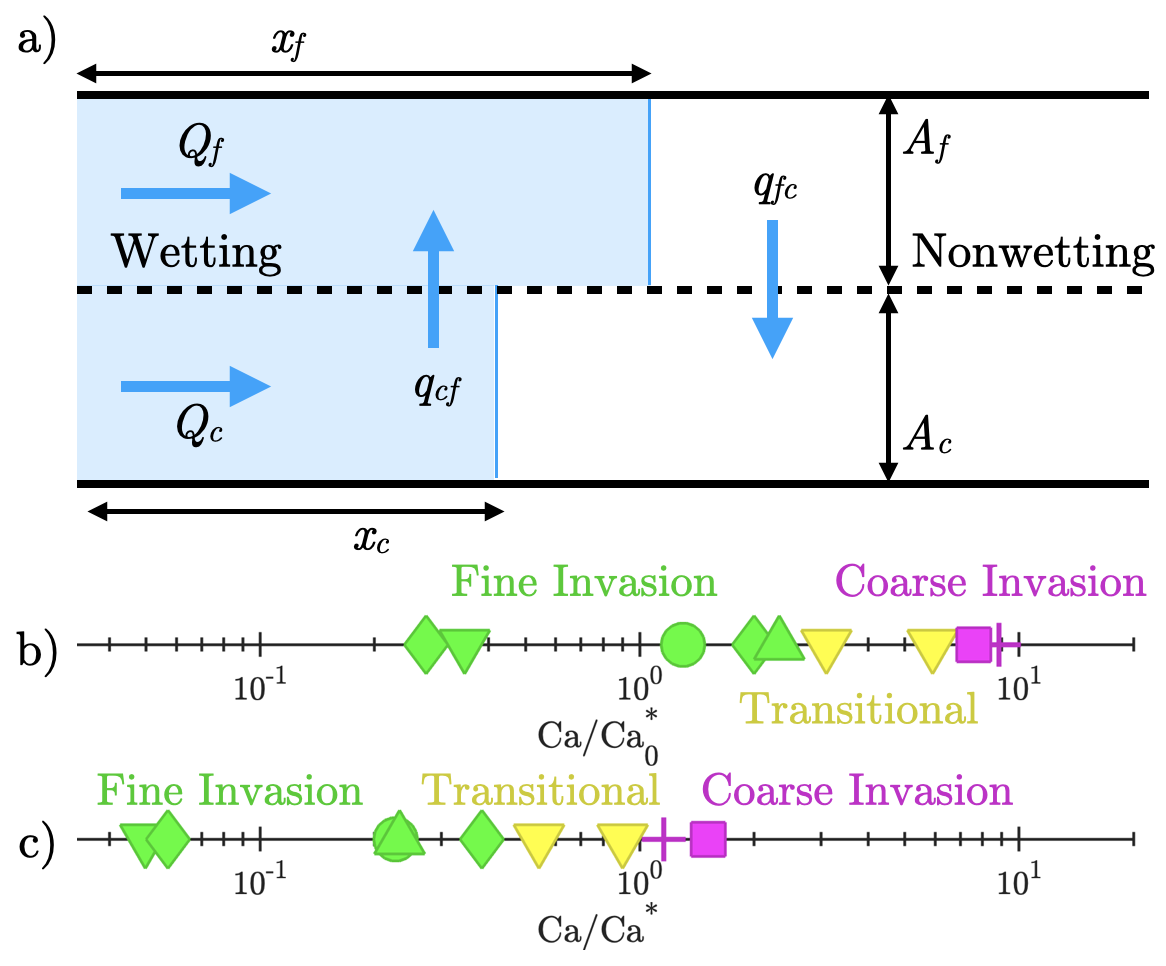}
    \caption{(a) Schematic of the theoretical model of imbibition in stratified porous media. We consider two parallel porous strata $i\in\{f,c\}$ oriented along the $x$ direction with equal length $\ell$, but with different pore throat sizes $a_{i}$, absolute permeabilities $k_{i}$, cross sectional areas $A_{i}$, flow rates $Q_{i}$, and extent of invasion by the wetting fluid $x_{i}$.  Crossflow upstream and downstream of the wetting/non-wetting fluid interface is represented by the flow rates per unit length along the $x$ direction $q_{cf}$ and $q_{fc}$.    (b) In the absence of crossflow, a linear stability analysis predicts a transition between invasion behaviors at $\text{Ca}=\text{Ca}_{0}^{*}$; the experiments exceed this prediction by a factor of $\sim 5$. (c) Incorporating crossflow, our analysis predicts a transition between invasion behaviors at $\text{Ca}=\text{Ca}^{*}$, where $\text{Ca}^{*}$ is given by Eq. \eqref{theory_eq}, in good agreement with the experimental data. Symbols are as described in the caption to Fig. \ref{primary_65x}.}
    \label{fig3}
\end{figure}

To understand this complex flow behavior, we analyze the distribution of pressures in each fluid and each stratum. First, we make the standard assumption that the flow is only along the length of the medium---strata cannot exchange fluid with each other in the transverse direction, as assumed in conventional ``pore doublet" models of heterogeneous porous media \cite{sorbie1995extended,chatzis1983dynamic,ridgway2002effect}. In each stratum, the invasion of the wetting fluid into the pores is guided by capillary suction; the pressure in the wetting fluid is reduced by the capillary pressure $p_{c,i}\sim\gamma/a_{i}$ compared to the non-wetting fluid at the interface between the two, where $a_{i}\approx0.08d_{i}$ is the pore throat radius \cite{lenormand1983mechanisms,princen1969capillary,princen1969capillary2,princen1970capillary,mason1986meniscus}. The pressure gradient in each fluid is described by Darcy's law, $\partial p_{i}/\partial x=-\mu_{j} q_{i}/k_{i}$, where $x$ represents the distance from the inlet, $j\in\{w,nw\}$ denotes the wetting or non-wetting fluid phase, and $q_{i}\equiv Q_{i}/A_{i}$ is the flux associated with the volumetric flow rate $Q_{i}$ in each stratum. The fluid pressure is equal across the strata at the inlet and outlet, and the sum of the flow rates is fixed throughout, $Q_{c}+Q_{f}=Q$. This model is sketched in Fig. \ref{fig3}a, which also indicates fluid crossflow between strata $q_{cf}$ and $q_{fc}$; however, in this initial version of the model, we set both of these terms to zero. 

To describe the invasion dynamics, we perform a linear stability analysis of this model at the incipient stage of imbibition, at which the wetting fluid begins to invade the medium from the inlet. Specifically, we perturb the position of the wetting/non-wetting fluid interface in the coarse stratum by a small amount and determine the conditions under which this perturbation grows or shrinks, as detailed further in the \textit{Appendix}. This analysis reveals that---similar to the experimental findings---the wetting fluid preferentially invades the fine stratum for $\text{Ca}<\text{Ca}^*_0$, and instead preferentially invades the coarse stratum for $\text{Ca}>\text{Ca}^*_0$, with the threshold value given by 
\begin{equation}
\text{Ca}^*_0=\frac{2\mu_w/\mu_{nw}}{\ell\left(1/k_{f}-1/k_{c}\right)}\left(\frac{1}{a_{f}} -\frac{1}{a_{c}} \right).
\label{eq1}
\end{equation}
\noindent This analysis quantifies the intuition provided by a scaling analysis of the underlying pressures \cite{zhou1997scaling}. In particular, at low flow rates, capillarity dominate the viscous effects, and we therefore expect the wetting fluid to imbibe faster into the fine stratum, for which the capillary suction $\propto 1/a_{i}$ is stronger. By contrast, at larger flow rates, viscous effects become comparable and eventually dominate capillarity, and we therefore expect the wetting fluid to imbibe faster into the coarse stratum, for which the permeability $k_{i}\propto a_{i}^{2}$ is larger and thus the viscous pressure drop $\propto 1/a_{i}^{2}$ is lower. We use the experimental measurements performed at varying Ca, $A_{c}/A_{f}$, and $\ell$ to directly test this prediction, as shown by the different symbols in Fig. \ref{fig3}b. Despite the qualitative agreement between the predictions of the linear stability analysis and the experimental observations, we do not find quantitative agreement: Eq. \eqref{eq1} underpredicts the transition by a factor of $\sim5$, as shown by the transition between preferential invasion of the fine stratum (green points) or the coarse stratum (magenta points) in Fig. \ref{fig3}b, suggesting that the model, which lacks crossflow, is incomplete. 

%%%%%%%%%%%%%%% PRIMARY IMBIBITION WITH CROSSFLOW
\begin{figure*}[htp!]
    \centering
    \includegraphics[width=0.95\linewidth]{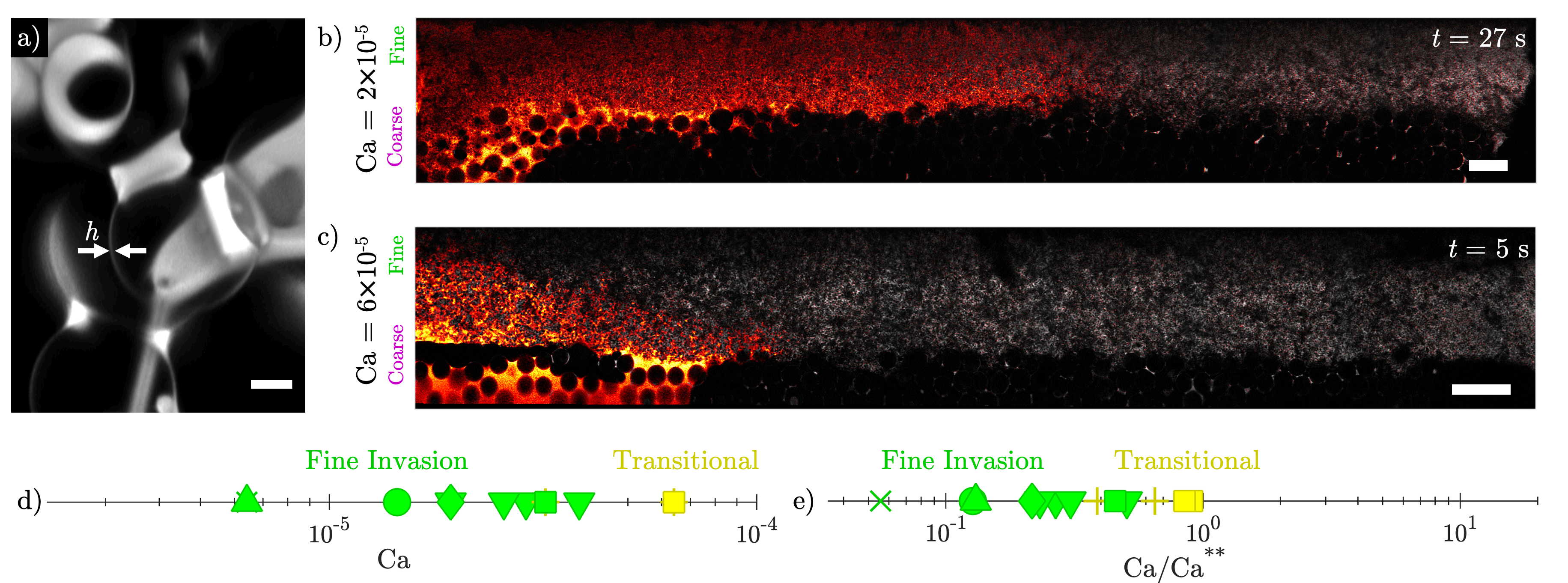}
        \caption{Wetting films alter invasion behavior during secondary imbibition. (a) 3D reconstruction of residual wetting films, shown in white, in a fine stratum. The beads are rendered transparent for ease of visualization; the wetting fluid surrounds non-wetting fluid, also rendered transparent, that fills the rest of the pore space. The residual wetting fluid forms thick pendular rings at the contacts between beads and thin films of thickness $h$ coat the rest of the bead surfaces. Scale bar denotes 50 $\mu$m. (b) and (c) show optical sections through the entirety of two stratified porous media, obtained a time $t$ after the wetting fluid begins to invade each medium. The red region shows the invading wetting fluid, the white shows pre-existing residual wetting fluid, and the black shows the undyed non-wetting fluid in the rest of the pore space. Imposed flow direction is from left to right. At a low capillary number Ca, wetting fluid preferentially invades the fine stratum, as shown in (b), while at a sufficiently large Ca, the wetting fluid instead invades both strata, as shown in (c). Scale bars denote 1 mm. (d) Summary of secondary imbibition experiments performed at different Ca; green points indicate preferential invasion into the fine stratum, while yellow points indicates that both strata are invaded at similar speeds. (e) Our analysis predicts a transition between invasion behaviors at $\text{Ca}=\text{Ca}^{**}$, where $\text{Ca}^{**}$ is calculated as described in the main text, in good agreement with the experimental data. Symbols are as described in the caption to Fig. \ref{primary_65x}, with the addition of the $\times$ indicating $ A_c/A_f \geq 1.5$ and $1.4\leq \ell < 2.0$ cm. }
    \label{secondary}
\end{figure*}
% a) Fluid enters the medium at time point 27 (140.60 seconds). It reaches the end of the fine stratum at time point 35 (183.76 seconds), i.e. velocity is 0.432 mm/s. Reaches end of coarse stratum at time point 39 (205.09 seconds), so the velocity is 0.289 mm/s. The length of the medium is 18.64 mm. 
% b) Fluid enters the medium at time point 27 (166.47 seconds). Time point 30 (185.24 seconds) is when the medium reaches the end of the coarse/fine stratum. The length of the medium is 24.735 mm. Superficial velocity is 1.32 mm/s

Previous theoretical work suggests that crossflow between strata can impact immiscible fluid displacement \cite{zapata1981theoretical,zhou1997scaling}. Inspired by this work, we incorporate crossflow in the model, both upstream and downstream of the wetting/non-wetting fluid interface, as indicated in Fig. \ref{fig3}a. Specifically, we introduce the \textit{ansatz} that a difference in pressures in the strata at a given longitudinal position $x$ drives a transverse crossflow between the strata: $q_{ij}(x) = \alpha_{ij} (p_i(x) - p_j(x))$, where $q_{ij}$ is the transverse volumetric flow rate from stratum $i$ to stratum $j$ per unit length along the medium, and $\alpha_{ij}$ is a proportionality constant. Performing a linear stability analysis of this extended model at the incipient stage of imbibition (\textit{Appendix}) again yields the prediction that the wetting fluid preferentially invades the fine stratum for $\text{Ca}<\text{Ca}^*$, and instead preferentially invades the coarse stratum for $\text{Ca}>\text{Ca}^*$, with the threshold value now given by 
\begin{multline}
\text{Ca}^*=\text{Ca}^*_0 +\\ \frac{\mu_w}{\gamma\ell\left(1/k_{f}-1/k_{c}\right)}\left(\frac{1}{k_fA_f}+\frac{1}{k_cA_c}\right)\int_{0}^{\ell}\int_{0}^{\xi}q_{fc}(x)~dx~d\xi,
\label{theory_eq}
\end{multline}
\noindent with
\begin{equation}
q_{fc}(x)=   \frac{\alpha_{fc}\left(p_{c,f} - p_{c,c}\right) \left(1-x/\ell\right)}{1 + \alpha_{fc} \left({\mu_{nw}}/{2}\right) \left( \frac{1}{k_c A_c} + \frac{1}{k_f A_f} \right) x (\ell-x)} ;
\label{crossflow}
\end{equation}
\noindent here, $\alpha_{fc}\approx C/\mu_{nw}\left(k_{f}^{-1}+k_{c}^{-1}\right)$ and $C$ is a constant that we take to be equal to 0.5. Notably, the crossflow term in Eq.~\eqref{crossflow} is positive, indicating that crossflow increases the threshold capillary number and suppresses invasion of the coarse stratum.  This is due to the fact that crossflow here moves the non-wetting fluid from the fine stratum to the coarse stratum, reducing the overall viscous dissipation; therefore, one has to increase Ca even further for viscous effects to become comparable to capillarity. A detailed analysis of how $\text{Ca}^*$ depends on the geometrical characteristics of the strata is given in the \textit{Appendix}. We use our experimental measurements to again test this prediction. Remarkably, we find excellent agreement between the experimental results and Eq. \eqref{crossflow}, with the transition between invasion behaviors arising at $\approx \text{Ca}^*$, as shown in Fig. \ref{fig3}c. This close agreement demonstrates the validity of our theory, and indicates that crossflow between strata impacts imbibition in stratified porous media.

\section{Secondary Imbibition}

%%%%%%%%%%%%%%% SECONDARY IMBIBITION

Many key processes, such as oil/gas recovery, groundwater remediation, and geological CO$_{2}$ sequestration, involve secondary imbibition---imbibition in a medium with pre-existing residual wetting films \cite{mohanty1987,blunt1995pore,blunt2002detailed,diaz1987simulation,idowu2010pore,mogensen1998dynamic,blunt1997pore,blunt1998physically,blunt1992simulation,constantinides2000effects,oren1998extending,patzek2000verification,ramakrishnan1986two}. Such films are thought to impact the invasion dynamics, although investigations of this behavior are woefully lacking; scattering of light from the film surfaces typically precludes direct characterization of their structure and their impact on the flow. Our experimental platform overcomes this challenge due to the close match between the refractive indices of the wetting fluid, non-wetting fluid, and glass beads. To generate residual wetting films, we inject the non-wetting fluid at a large flow rate $Q_{nw}=500$ mL/h into media that are initially saturated with the wetting fluid. At this large flow rate, the non-wetting fluid invades the pore space of both strata \cite{datta2013drainage}, within which it is surrounded by residual films of the wetting fluid that remain trapped in the crevices and along the solid surfaces of the medium  \cite{lenormand1988numerical,blunt1995pore,blunt2002detailed,diaz1987simulation,hughes2000pore,idowu2010pore,mogensen1998dynamic,blunt1997pore,blunt1998physically,blunt1992simulation,constantinides2000effects,oren1998extending,patzek2000verification,ramakrishnan1986two}. A magnified 3D reconstruction is shown in Fig. \ref{secondary}a: thick pendular rings are formed at the contacts between beads, while thinner films of thickness $h$ coat the surfaces of the beads. 

We then commence secondary imbibition by re-injecting the wetting fluid into the medium. Again, we observe different invasion behaviors at different Ca: at small Ca, the fluid preferentially invades the fine stratum, as exemplified by Fig. \ref{secondary}b, while at sufficiently large Ca, we find that both strata are invaded at comparable speeds, as exemplified by Fig. \ref{secondary}c. However, unlike in primary imbibition, we do not observe preferential invasion into the coarse stratum at the largest Ca tested. Instead, as summarized in Fig. \ref{secondary}d, the presence of residual wetting films extends the transition between invasion behaviors by a factor of $\approx2.5$ to $\text{Ca}\gtrsim5\times10^{-5}$.

To describe this behavior, we examine the alteration to the flow dynamics caused by the residual wetting fluid. The hydraulic properties of a bead packing are thought to be controlled primarily by the geometry of the pore throats, the narrow polygonal constrictions between the surfaces of adjacent beads \cite{bear2013dynamics}. We therefore focus on the alteration caused by the thin wetting films that coat the bead surfaces and reduce the inscribed radii $a$ of the pore throats. In particular, we use a standard capillary bundle model of the pore throats (\textit{Appendix}) to estimate the modified non-wetting fluid permeability downstream of the wetting/non-wetting fluid interface, and the modified capillary pressure within each stratum $i$, as $k'_{i}\approx k_{i}\left(\mu_{nw}/\mu_{w}\right)\left[2-\left(2-\mu_{w}/\mu_{nw}\right)\left(1-h_{i}/a_i\right)^2\right]$ and $p'_{c,i}\sim\gamma/\left(a_{i}-h_{i}\right)$, respectively. Applying these modified parameters in Eqs. \eqref{theory_eq}--\eqref{crossflow} thus yields the threshold value of Ca for secondary imbibition, which we denote Ca$^{**}$. Our 3D visualization indicates that $h_{i}\lesssim600$ nm, the diffraction limit of our imaging. Consistent with this measurement, we find that the transition between invasion behaviors arises at $\approx\text{Ca}^{**}$ with $h\sim100$ nm, as shown in Fig. \ref{secondary}d; this value of $h$ is in reasonable agreement with previous theoretical analysis of the interplay between capillary stresses, viscous stresses, and disjoining pressure in the formation of these films \cite{scriven88,scriven89}. Thus, the theoretical framework presented here can capture the heterogeneous dynamics of both primary and secondary imbibition. Unraveling how the wetting film thickness depends on flow behavior and the physico-chemical interactions between the fluids and the medium will be a useful direction for future work. 

%\textcolor{red}{, as previously conjectured \cite{zhou1997scaling}}

\section{Discussion}
%%%%%%%%%%%%%%% CONCLUSION
Our experiments reveal the heterogeneous invasion behaviors that can arise in stratified porous media during forced imbibition. Using direct visualization in 3D media, we show that this behavior is sensitive to the imposed Ca: at small Ca, the wetting fluid preferentially invades the fine stratum, while above a threshold Ca, the wetting fluid instead preferentially invades the coarse stratum. These findings thus complement our previous work investigating forced drainage, the converse process in which a non-wetting fluid displaces a wetting fluid, in a stratified porous medium \cite{datta2013drainage}, for which heterogeneous invasion behaviors also can arise. However, the pore-scale physics of drainage are fundamentally different from those underlying imbibition \cite{lenormand1990liquids}, leading to completely different fluid displacement behaviors and threshold $\textrm{Ca}^*$ marking the transition between these behaviors. 

Our experiments provide qualitative verification of a previous conjecture \cite{zhou1997scaling}, which was developed by analyzing the balance between capillary and viscous forces in an averaged multi-phase form of Darcy's law. However, this averaged treatment does not explicitly treat the capillary pressure difference across the sharp, moving interface between the wetting and non-wetting fluids, nor does it explicitly treat the viscous pressure gradients in each fluid and in each stratum. Hence, we construct a model of the flow that explicitly incorporates these physics. The linear stability analysis of this model presented in the \textit{Appendix} captures the transition between invasion behaviors when crossflow between the strata is considered. More detailed studies of this crossflow will be an interesting direction for future work. Importantly, our analysis quantifies how the threshold Ca depends on the permeabilities, pore sizes, and cross-sectional areas of the strata, the fluid viscosities and interfacial tension, the thickness of residual wetting films in the pore space, and the length of the overall medium. Our work therefore provides quantitative guidelines to predict and control the heterogeneous dynamics of imbibition in diverse stratified porous media over a range of length and time scales. Hence, we anticipate that our findings will help motivate both fundamental and applied work managing or even harnessing stratification in energy, environmental, and industrial applications.

For example, these results could provide insight into the pathways taken by fluids during oil/gas recovery from a stratified reservoir or from microporous media \cite{king2018microstructural,ingham2010geomaterials,tanino2012capillary,bear2013dynamics,galloway2012terrigenous}---an increasingly important application given society's rising energy demands. Another operationally similar application of our work will be to inform strategies for injecting fluid to recover non-aqueous contaminants from a stratified porous groundwater aquifer \cite{bear2013dynamics,rabideau1994}---another critical application due to the increasing need for water security. Models of these processes often treat a stratified reservoir or aquifer as a non-stratified, homogeneous medium using a strata-averaged permeability. Our work demonstrates that such averaging can lead to misleading results: for example, imbibition in a non-stratified, homogeneous medium with a strata-averaged permeability proceeds uniformly across the medium, while our work reveals that the wetting fluid instead preferentially enters certain strata, potentially leading to incomplete non-wetting fluid recovery from the medium, and necessarily reaching the outlet of the medium after a shorter time than in the homogeneous case. Thus, in some circumstances, averaged models may incorrectly predict both the fluid distribution and the rate of fluid displacement during oil/gas recovery from a stratified reservoir or during contaminant removal from a stratified aquifer. By contrast, our theoretical framework provides a way to more accurately predict these quantities, given the structural properties of the medium, fluid properties, and injection conditions.

Our results could also impact the use and design of building materials, which are often composed of multiple layers of materials with different structural properties and permeabilities \cite{hall,rahman2015recycled,takahashi1996acoustic}. These differences can strongly impact moisture retention, with potentially severe consequences. For example, our work suggests that moisture may seep into specific layers preferentially (e.g., during heavy rainfall), potentially leading to localized water damage, compromised structural integrity, and harm to health. Our findings may help guide efforts to mitigate or prevent such deleterious effects---for example, by indicating layers that would benefit from additional mechanical reinforcement or wettability treatment. Similarly, our work may help guide efforts to control moisture wicking in layered microporous fabrics \cite{elnashar2005volume, adams1991permeability, ahn1991simultaneous}---for example, by indicating how layers of different permeabilities can be designed to concentrate wicking in one area, potentially improving thermal comfort \cite{bivainyte2011investigation}. Finally, we anticipate that our results may find use in the emerging field of paper microfluidics \cite{castro2017characterizing, martinez2008three, conn2014visualizing,jang2015facile}, in which microporous cellulose-based materials are used to guide liquids through imbibition. In some cases, these materials are made of layers with different microstructures and permeabilities---heterogeneities that have been found to give rise to different amounts of fluid mixing. Our results help rationalize how such structural heterogeneities impact fluid distributions and mixing, and could be used to guide future efforts to direct fluid delivery in paper microfluidics by designing strata of different permeabilities.

\section{Appendix: Theoretical model of imbibition in stratified porous media}
\setcounter{equation}{0}
\renewcommand\theequation{S.\arabic{equation}}
To understand the complex flow behavior that arises during imbibition, we analyze the distribution of pressures in each fluid and each stratum, inspired by conventional ``pore doublet" models of heterogeneous porous media \cite{sorbie1995extended,chatzis1983dynamic,ridgway2002effect}, as sketched in Fig. \ref{fig3}a.\\

\textit{Mass conservation.} Because the strata are in contact with each other over their entire length, mass conservation implies that 
\begin{equation}
Q = Q_c + Q_f  
\label{eq:MassCons1}
\end{equation}
\begin{equation}
\mathrm{with}~Q_c=A_c \phi \frac{dx_c}{dt} + \int_0^{x_f} q_{cf}(x') ~dx'  
\label{eq:MassCons2}
\end{equation}
\begin{equation}
\mathrm{and}~Q_f=A_f \phi \frac{dx_f}{dt} - \int_0^{x_f} q_{cf}(x') ~dx';
\label{eq:MassCons3}
\end{equation}
%\begin{align}
%Q &= Q_c + Q_f, \nonumber \\
%A_c \phi_c \frac{dx_c}{dt} &= Q_c - \int_0^{x_f} q_{cf}(x') ~dx', \nonumber \\
%A_f \phi_f \frac{dx_f}{dt} &= Q_f + \int_0^{x_f} q_{cf}(x') ~dx.
%\label{eq:MassCons}
%\end{align}
here, $x_c(t)$ and $x_f(t)$ represent the longitudinal positions of the non-wetting/wetting fluid interface in the coarse and fine strata, respectively, and $\phi$ is the porosity, taken to be equal for the different strata because they are both dense packings of spherical beads. We assume hereafter that the interface in the coarse stratum is ahead of the interface in the fine stratum, although revising this derivation with the opposite assumption yields the same results. Further, we assume that crossflow only occurs within a given fluid, i.e., there is no crossflow across different fluid phases. \\ \\

\textit{Distribution of fluid pressures.} The pressure gradient in each stratum and fluid is described by Darcy's law, 
\begin{equation}
\frac{\partial p_{i}}{\partial x}=-\frac{\mu_{j} Q_{i}}{k_{i}A_{i}},
\label{eq:Darcy}
\end{equation}
where $x$ represents the distance from the inlet, $i\in\{c,f\}$ denotes the stratum, $j\in\{w,nw\}$ denotes the fluid phase, and $Q_{i}$ is the volumetric flow rate in each stratum of cross-sectional area $A_{i}$. The flow rate can vary along the length of each stratum due to crossflow. Further, at the interface between the fluids, the pressure in the wetting fluid is reduced by the capillary pressure $p_{c,i}\equiv2\gamma/a_{i}$ compared to the non-wetting fluid \cite{lenormand1983mechanisms,princen1969capillary,princen1969capillary2,princen1970capillary,mason1986meniscus}.

Integrating Darcy's law in the coarse stratum from the inlet to the outlet results in an expression for the total pressure drop across the medium:
\begin{equation}
\begin{split}
& p_{inlet} - p_{outlet} = \\ 
& \int_0^{x_f} \frac{\mu_{w}}{k_c} \left( \frac{Q_c - \int_0^x q_{cf}(x')~dx'}{A_c}\right) ~dx   \\ & +\int_{x_f}^{x_c} \frac{\mu_{w}}{k_c} \left( \frac{Q_c - \int_0^{x_f} q_{cf}(x')~dx'}{A_c}\right) ~dx
\\ & + \int_{x_c}^{\ell} \frac{\mu_{nw}}{k_c} \left( \frac{Q_c - \int_0^{x_f} q_{cf}(x')~dx' + \int_{x_c}^{x} q_{fc}(x')~dx'}{A_c}\right) ~dx\\ &   - p_{c,c}.
\end{split}
\label{eq:Darcy1}
\end{equation}
\noindent Repeating this procedure for the fine stratum results in another expression for the total pressure drop across the medium:
\begin{equation}
\begin{split}
& p_{inlet} - p_{outlet} = \\ 
& \int_0^{x_f} \frac{\mu_{w}}{k_f} \left( \frac{Q_f + \int_0^x q_{cf}(x')~dx'}{A_f}\right) ~dx \\ &+   \int_{x_f}^{x_c} \frac{\mu_{nw}}{k_f} \left( \frac{Q_f + \int_0^{x_f} q_{cf}(x')~dx'}{A_f}\right) ~dx
\\ & + \int_{x_c}^{\ell} \frac{\mu_{nw}}{k_f} \left( \frac{Q_f + \int_0^{x_f} q_{cf}(x')~dx' - \int_{x_c}^{x} q_{fc}(x')~dx'}{A_f}\right) ~dx\\ &  - p_{c,f}.
\end{split}
\label{eq:Darcy2}
\end{equation}
\noindent Because the fluid pressure is equal across the strata at the inlet and outlet, we can subtract Eq.~\eqref{eq:Darcy1} from \eqref{eq:Darcy2}:
\begin{equation}
\begin{split}
& \mu_{w} \int_0^{x_f} \bigg[\left( \frac{Q_f + \int_0^x q_{cf}(x')~dx'}{k_f A_f}\right) \\ &- \left( \frac{Q_c - \int_0^x q_{cf}(x')~dx'}{k_c A_c}\right) \bigg] ~dx~+ 
\\ & \mu_{nw} \int_{x_f}^{x_c} \bigg[\left( \frac{Q_f + \int_0^{x_f} q_{cf}(x')~dx'}{k_f A_f}\right)\\ & -  \frac{\mu_{w}}{\mu_{nw}} \left( \frac{Q_c - \int_0^{x_f} q_{cf}(x')~dx'}{k_c A_c}\right)\bigg] ~dx~+
\\ & \mu_{nw} \int_{x_c}^{\ell} \bigg[\left( \frac{Q_f + \int_0^{x_f} q_{cf}(x')~dx' - \int_{x_c}^{x} q_{fc}(x')~dx'}{k_f A_f}\right) 
 \\ & - \left( \frac{Q_c- \int_0^{x_f} q_{cf}(x')~dx' + \int_{x_c}^{x} q_{fc}(x')~dx'}{k_c A_c}\right) \bigg] ~dx  
\\ & -  \left( p_{c,f} - p_{c,c} \right)= 0.
\end{split}
\label{eq:entirechannel}
\end{equation}\\

\textit{Crossflow between the strata.} Motivated by Darcy's law, we introduce the \textit{ansatz} that a difference in pressures in the strata at a given longitudinal position $x$ drives a transverse crossflow between the strata: $q_{ij}(x) = \alpha_{ij} (p_i(x) - p_j(x))$, where $\alpha_{ij}$ is a proportionality constant. Specifically, a pressure difference $p_i(x) - p_j(x)$ across the strata drives a transverse volumetric flow rate $Q_{ij}(x)$ from stratum $i$ to $j$; thus, the transverse volumetric flow rate per unit length along the medium $q_{ij}(x)\equiv Q_{ij}(x)/\Delta x$, where $\Delta x$ is a fixed differential length along the imposed flow direction. Applying Darcy's law to describe this transverse flow in each stratum yields $p_i(x) - p_j(x)=\frac{\mu q_{ij}(x)}{h}\left(\frac{\ell_{i}}{k_{i}}+\frac{\ell_{j}}{k_{j}}\right)$, where $\mu$ is the fluid viscosity, $h$ is the lateral width of the medium, and $\{\ell_{i},\ell_{j}\}$ and $\{k_{i},k_{j}\}$ are the crossflow lengths and the permeabilities of stratum $i$ or $j$, respectively. Thus, $\alpha_{ij}=\frac{h}{\mu\left(\ell_{i}/k_{i}+\ell_{j}/k_{j}\right)}$. Finally, assuming for simplicity that $\ell_{i}$ and $\ell_{j}$ can be approximated by a characteristic crossflow length scale $\approx h/C$ yields the final estimate of $\alpha_{ij}\approx C/\mu\left(k_{f}^{-1}+k_{c}^{-1}\right)$, where $C=0.5$ acts as an empirical fitting parameter. Future extensions of our work could explicitly derive and incorporate the specific values of $\ell_{i}$ and $\ell_{j}$, which would improve the accuracy of our prediction and potentially circumvent the use of $C$ as an empirical parameter. 

To obtain the crossflow, we then integrate Darcy's law up to some arbitrary distance $x$ and simply subtract the pressures obtained in each stratum at that location. This procedure yields the following integral equations for the crossflow in the wetting and non-wetting fluids, respectively, where $\alpha_{cf}$ and $\alpha_{fc}$ reflect the different directions of crossflow in the different fluid phases, as indicated in Fig. \ref{fig3}a:
\begin{equation}
\begin{split}
&\frac{q_{cf}}{\alpha_{cf} \mu_{w}} - \left( \frac{1}{k_c A_c} +  \frac{1}{k_f A_f} \right) \int_0^{x} \left( \int_0^\xi q_{cf}(x')~dx' \right)~d\xi\\ &  =  \left( \frac{Q_f}{k_f A_f} - \frac{Q_c}{k_c A_c} \right) x,
\end{split}
\label{eq:crossflowflow1}
\end{equation}
\begin{equation}
\begin{split}
& \frac{q_{fc}}{\alpha_{fc}} = 
\\& -\mu_{w} \int_0^{x_f} \Bigg[\left( \frac{Q_f + \int_0^\xi q_{cf}(x')~dx'}{k_f A_f}\right) \\&- \left( \frac{Q_c - \int_0^\xi q_{ab}(x')~dx'}{k_c A_c}\right) \Bigg] ~d\xi~ 
\\ & -\mu_{nw} \int_{z_f}^{z_c} \Bigg[\left( \frac{Q_f + \int_0^{x_f} q_{cf}(x')~dx'}{k_f A_f}\right) \\&-  \frac{\mu_{w}}{\mu_{nw}} \left( \frac{Q_c - \int_0^{x_f} q_{cf}(x')~dx'}{k_c A_c}\right)\Bigg] ~d\xi~
\\ & -\mu_{nw} \int_{z_c}^{z} \Bigg[\left( \frac{Q_f + \int_0^{x_f} q_{cf}(x')~dx' - \int_{z_c}^{\xi} q_{fc}(x')~dx'}{k_f A_f}\right) 
 \\ &- \left( \frac{Q_c- \int_0^{x_f} q_{cf}(x')~dx' + \int_{z_c}^{\xi} q_{fc}(x')~dx'}{k_c A_c}\right) \Bigg] ~d\xi  
\\ & +  \left( p_{c,f} - p_{c,c} \right).
\end{split}
\label{eq:crossflowflow2}
\end{equation}\\

\textit{Linear stability analysis.} Solving the six coupled integral equations Eqs.~\eqref{eq:MassCons1}, \eqref{eq:MassCons2}, \eqref{eq:MassCons3}, \eqref{eq:entirechannel}, \eqref{eq:crossflowflow1}, \eqref{eq:crossflowflow2} would provide a full description of the evolution of the non-wetting/wetting fluid interface position in each of the two strata. At low flow rates, capillarity dominate the viscous effects, and we therefore expect the wetting fluid to imbibe faster into the fine stratum, for which the capillary suction $\propto 1/a_{i}$ is stronger. By contrast, at larger flow rates, viscous effects become comparable and eventually dominate capillarity, and we therefore expect the wetting fluid to imbibe faster into the coarse stratum, for which the permeability $k_{i}\propto a_{i}^{2}$ is larger and thus the viscous pressure drop $\propto 1/a_{i}^{2}$ is lower. 

To test this hypothesis, we use a linear stability analysis to probe the dynamics of this system \cite{Protiere10,Housseiny14}. Specifically, we assume that the non-wetting/wetting fluid interfaces in the two strata are at the same location, $x_c = x_f$, and then perturb the interface position in the coarse stratum by a small amount, $x_c = x_f + \epsilon$, where $\epsilon/\ell \ll 1$. We then probe the subsequent motion of these two interfaces to determine whether this initial infinitesimal perturbation grows or decays in time: a growing perturbation implies that the wetting fluid would preferentially imbibe into the coarse stratum, whereas a decaying perturbation indicates stable propagation together. The boundary between these two scenarios therefore determines the threshold flow rate. 

Specifically, we rearrange the mass conservation Eqs.~\eqref{eq:MassCons1}, \eqref{eq:MassCons2}, \eqref{eq:MassCons3} to express the interface speed in terms of the perturbation $\epsilon$:
\begin{align}
\frac{d x_c}{d t} & = \frac{Q/\phi + A_f  \frac{d \epsilon}{dt}}{A_c + A_f }, \\
\frac{d x_f}{d t} &= \frac{Q/\phi - A_c \frac{d \epsilon}{dt}}{A_c + A_f }.
\label{eq:ConsMass2}
\end{align} 

\noindent At the beginning of the experiment both interfaces are near the inlet, i.e., $x_f \to 0$, and $x_c \to \epsilon$. We can therefore neglect the crossflow terms in the invading wetting fluid, and simply relate the flow rate in each stratum to the interfacial motion at the onset, i.e., $Q_c = A_c \phi\left(dx_c/dt\right)$ and $Q_f = A_f \phi \left(dx_f/dt\right)$. Substituting these back into Eq.~\eqref{eq:entirechannel}, we then arrive at the following expression for the growth of a perturbation:

 \begin{align}
 \begin{split}
&\beta \frac{d \epsilon}{d t} = \left( p_{c,c} - p_{c,f} \right) + \frac{\mu_{nw} Q\ell }{A} \left( \frac{1}{k_f} - \frac{1}{k_c} \right)\\& - \mu_{nw} \left(\frac{1}{k_c A_c} + \frac{1}{k_f A_f} \right) \int_0^\ell \int_0^\xi q_{fc}(x)~dx~d\xi,
\end{split}
\label{eq:depsilondt}
\end{align}
where $\beta \equiv \frac{\mu_{nw} \phi\ell}{A}  \left( \frac{A_c}{k_f} + \frac{A_f}{k_c} \right) > 0$. The first term on the right represents the contribution of capillarity, the second term represents the viscous flow contribution, and the last term shows the influence of the crossflow. To determine whether the perturbations grow or decay, we then need to know the sign of $d \epsilon / dt$, and therefore the sign of the right-hand side of Eq.~\eqref{eq:depsilondt}. \\

\textit{Absence of crossflow.} In the absence of crossflow, $q_{fc}=0$, simplifying the analysis of Eq.~\eqref{eq:depsilondt}. In this case, for perturbations to grow, i.e., for the wetting liquid to imbibe the coarse stratum faster, we need to have $\left( p_{c,c} - p_{c,f} \right) + \frac{\mu_{nw} Q\ell }{A} \left( \frac{1}{k_f} - \frac{1}{k_c} \right)>0$, which can be expressed as $\textrm{Ca} > \textrm{Ca}_{0}^*$, where the threshold capillary number is given by:
\begin{equation}
\textrm{Ca}_{0}^* = \frac{\left(\mu_{w}/\mu_{nw}\right) \left(p_{c,f} - p_{c,c} \right)}{\gamma \ell \left( 1/k_f - 1/k_c \right)},
\label{eq:CacrNoflow}
\end{equation}
which is equivalent to Eq. \eqref{eq1}. This expression reflects the arguments above regarding the competition between viscous and capillary effects, which is ultimately modulated by the geometric features of the medium, i.e., the differences in the permeabilities and pore sizes of the strata. \\

\textit{Including crossflow.} Equation \eqref{eq:crossflowflow2} is an integral equation for the crossflow in the non-wetting fluid. Using the conservation of mass Eq.~\eqref{eq:ConsMass2}, and taking the limit of $x_f \to 0$ and $x_c \to \epsilon$ as done above yields a simplified form of Eq.~\eqref{eq:crossflowflow2}:
\begin{equation}
\begin{split}
&\frac{q_{fc}(x)}{\alpha_{fc}} =  \beta \left(\frac{x}{\ell}\right) \frac{d \epsilon}{d t} + \frac{\mu_{nw} Qx}{A}  \left( \frac{1}{k_c} - \frac{1}{k_f} \right) \\&+ \left(p_{c,f} - p_{c,c} \right) + \mu_{nw} \left(\frac{1}{k_c A_c} + \frac{1}{k_f A_f}  \right) \int_0^x\int_0^\xi q_{fc}~dx'~d\xi.
\end{split}
\label{eq:crossflowflow2simp}
\end{equation}

\noindent Substituting for the $d\epsilon/dt$ term from Eq.~\eqref{eq:depsilondt} into Eq.~\eqref{eq:crossflowflow2simp} yields a more simplified explicit equation for the crossflow term:
\begin{equation}
\begin{split}
&\frac{q_{fc}(x)}{\alpha_{fc}} = \left(p_{c,f} - p_{c,c} \right) \left( 1 - \frac{x}{\ell} \right) \\&+ \mu_{nw} \left(\frac{1}{k_c A_c} + \frac{1}{k_f A_f} \right) \bigg( \int_0^x \int_0^\xi q_{fc}~dx'~d\xi \\&- \frac{x}{\ell} \int_0^\ell \int_0^\xi q_{fc}~dx'~d\xi  \bigg),
\end{split}
\label{eq:crossflowflow2simpf}
\end{equation}
which can be solved analytically to obtain the following expression for the crossflow term:
\begin{equation}
q_{fc}(x)=   \frac{\alpha_{fc}\left(p_{c,f} - p_{c,c}\right) \left(1-x/\ell\right)}{1 + \alpha_{fc} \left({\mu_{nw}}/{2}\right) \left( \frac{1}{k_c A_c} + \frac{1}{k_f A_f} \right) x (\ell-x)} .
\label{eq:crossflowflow2simpfinal}
\end{equation}
In the limit of $x \to 0$, the crossflow term further simplifies to $q_{fc} = \alpha_{fc} \left( p_{c,f} - p_{c,c} \right)$, which indicates that the pressure difference between the two strata right next to the interfaces is due to the Laplace pressure difference.  In the other limit of $x \to \ell$, the crossflow term vanishes, and this is consistent with the fact that both strata experience the same pressure at the outlet. 
\begin{figure}
  \centerline{\includegraphics[width=0.5\textwidth]{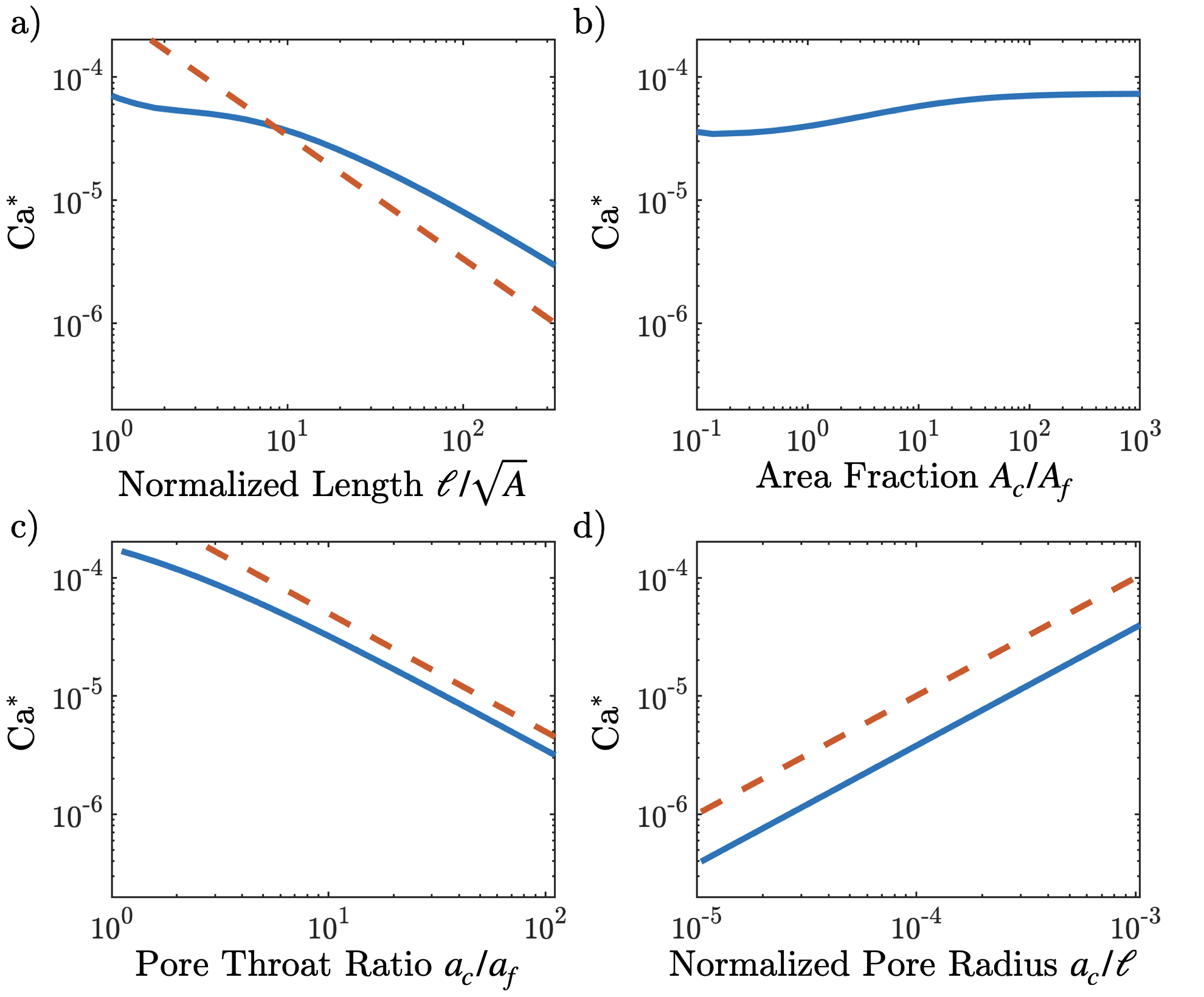}}% Images in 100% size
%  \centerline{\includegraphics[natheight=7cm,natwidth=13cm]{Drop.eps}}
\caption{Dependence of the predicted threshold $\textrm{Ca}^*$ on the geometric characteristics of the strata, as shown by the blue curves. All parameters except the one being varied are held fixed using values representative of our experiments. Orange dashed lines show comparison to asymptotic power-law scaling behaviors. (a) Dependence on the length of the medium, $\ell$, normalized by the characteristic transverse dimension $\sqrt{A}$. For small $\ell$, $\textrm{Ca}^*$ only decreases weakly with increasing $\ell$, eventually asymptoting to the $\sim\ell^{-1}$ scaling shown by the orange dashed line. Our experiments only probe $\ell/\sqrt{A}$ varying in a narrow range from $\sim5$ to $17$, for which the predicted $\textrm{Ca}^*$ does not appreciably vary and is in good agreement with the experimentally-determined threshold. (b) $\textrm{Ca}^*$ only varies slightly with the area fraction $A_{c}/A_{f}$. Thus, despite our experiments having $A_{c}/A_{f}$ varying from $\sim0.2$ to $4$, the predicted $\textrm{Ca}^*$ does not appreciably vary and is in good agreement with the experimentally-determined threshold. (c) $\textrm{Ca}^*$ decreases with increasing pore throat ratio $a_{c}/a_{f}$, eventually asymptoting to the $\sim(a_{c}/a_{f})^{-1}$ scaling shown by the orange dashed line. (d) Keeping the pore throat ratio $a_{c}/a_{f}$ fixed, but proportionally increasing both $a_{c}$ and $a_{f}$, increases $\textrm{Ca}^*$ linearly, as shown by the comparison with the $\sim a_{c}$ scaling shown by the orange dashed line.}
\label{fig:geometricdependence}
\end{figure}

\begin{figure}
  \centerline{\includegraphics[width=0.4\textwidth]{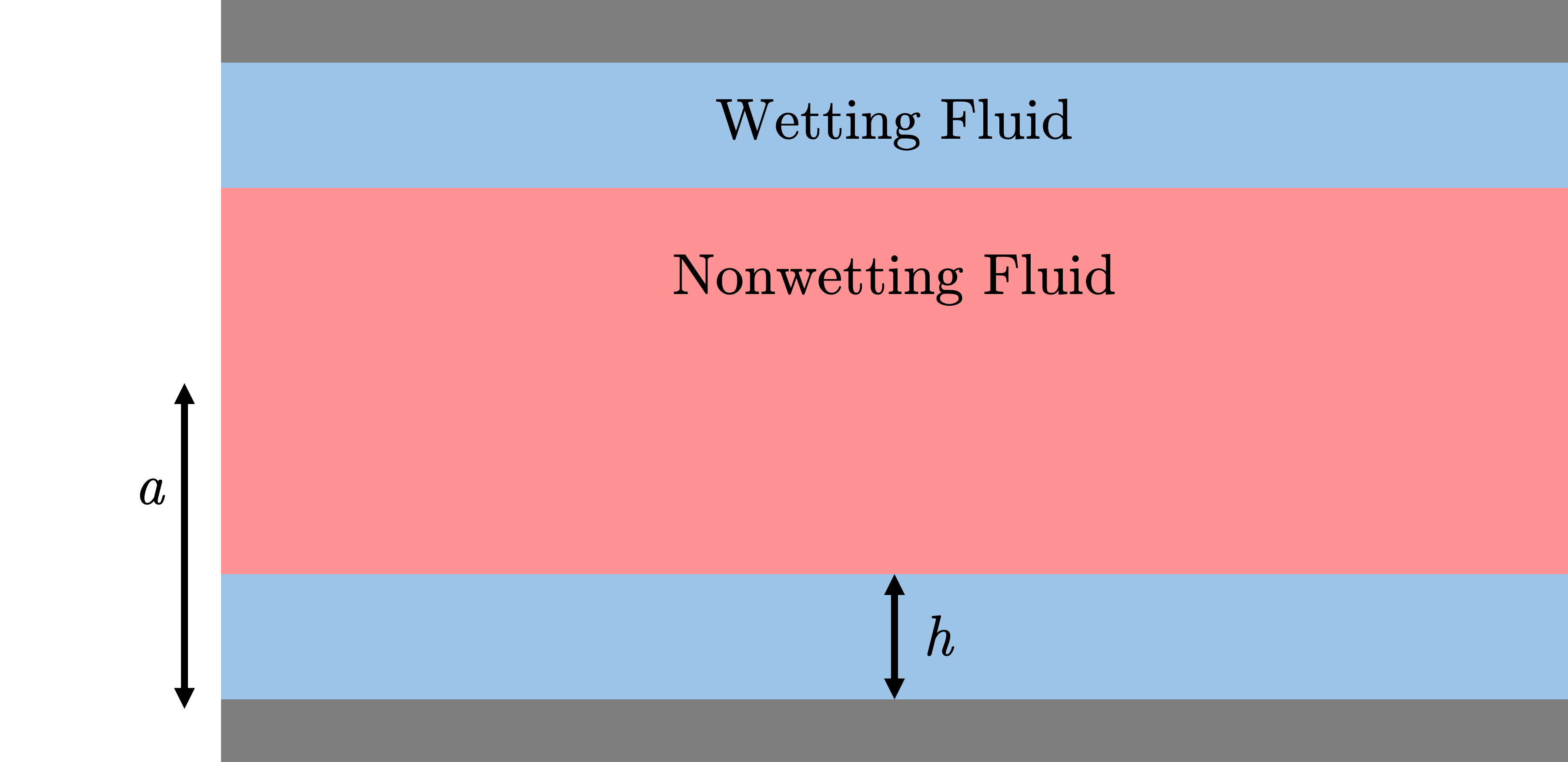}}% Images in 100% size
%  \centerline{\includegraphics[natheight=7cm,natwidth=13cm]{Drop.eps}}
\caption{Schematics of a cylindrical pore throat covered with thin films of the wetting fluid. $a$ represents the initial radius of the pore throat, $h$ represents the thickness of the thin wetting films, and $a-h$ represents the radius of the nonwetting fluid.}
\label{fig:secondaryimbibition}
\end{figure}

We can now substitute the expression for the crossflow, Eq.~\eqref{eq:crossflowflow2simpfinal}, back into the right-hand side of Eq.~\eqref{eq:depsilondt} to obtain the threshold capillary number:
\begin{equation}
\begin{split}
&\textrm{Ca}^* = \frac{\left(\mu_{w}/\mu_{nw}\right)}{\gamma l \left( 1/k_f - 1/k_c \right)} \bigg[ \left(p_{c,f} - p_{c,c} \right) \\&+ \mu_{nw} \left(\frac{1}{k_c A_c} + \frac{1}{k_f A_f} \right) \int_0^\ell \int_0^\xi q_{fc}(x)~dx~d\xi\ \bigg],
\end{split}
\label{eq:Cacrflow}
\end{equation}
which is equivalent to Eq. \eqref{theory_eq}. Since the crossflow term in Eq.~\eqref{eq:crossflowflow2simpfinal} is positive, it is therefore clear that presence of crossflow increases the threshold capillary number and suppresses invasion of the coarse stratum. This is due to the fact that crossflow here moves the fluid from the fine stratum to the coarse stratum, reducing the viscous dissipation. Therefore, one has to increase the flow rate even further to make viscous forces become comparable to capillary forces, causing the invading fluid to preferentially imbibe into the coarse stratum. To aid with general application of our results, we highlight the dependencies of the threshold $\textrm{Ca}^*$ on the different geometric characteristic of the medium in Fig.~\ref{fig:geometricdependence}. Exploring these dependencies over a broad range of the geometric parameters will be a useful direction for future work.\\

\textit{Primary vs secondary imbibition.} Our imaging reveals the structure of the residual wetting fluid after non-wetting fluid injection, prior to secondary imbibition: we observe thick pendular rings at the contacts between beads, with thinner films of thickness $h$ coating the surfaces of the beads in between. Because the hydraulic properties of a bead packing are thought to be controlled primarily by the geometry of the pore throats, the narrow polygonal constrictions between the surfaces of adjacent beads \cite{bear2013dynamics}, we focus on the alteration to the flow dynamics caused by the thin wetting films that coat the bead surfaces and reduce the inscribed radii $a$ of the pore throats. This pore throat reduction alters two key features of each stratum: the capillary pressure and the permeability. Because $\cos\theta\approx1$, where $\theta\approx0$ is the contact angle, the modified capillary pressure within each stratum $i$ can be approximated as $p'_{c,i}\equiv2\gamma/\left(a_{i}-h\right)$. To estimate the modified permeability of each stratum, we use a standard capillary bundle model of the pore throats, as described further below.

Consider a capillary tube, representing a pore, coated with thin films of the wetting fluid as shown in Fig.~\ref{fig:secondaryimbibition}. In the absence of these thin films, it is straightforward to solve the Stokes equations, which lead to a Poiseuille flow of non-wetting fluid through the tube: $dp/dx = -8 \mu_{nw} Q_{nw} / (\pi a^4)$, where $Q_{nw}$ is the volumetric flow rate of non-wetting fluid through the tube and $\mu_{nw}$ is the non-wetting fluid dynamic shear viscosity. Using Darcy's law as given in Eq.~\eqref{eq:Darcy}, we can then write the absolute permeability of the cylindrical tube as $k=a^2 / 8$. 

In the presence of thin films, we need to solve the Stokes equations in both fluids, and impose continuity of velocity and shear stress on the interface between the two fluids. Doing so then leads to the following relationship between the pressure drop and flow rate of the non-wetting fluid: $dp/dx = - \mu_{nw} Q_{nw} / (\pi (a-h)^2 k')$, where $Q_{nw} = \int_{0}^{a-h} u_{nw} (2 \pi r)~dr$,  $u_{nw} = \frac{1}{4 \mu_w \mu_{nw}} \frac{dp}{dx} \left[ -\mu_{nw} a^2 + \mu_w r^2 + (a-h)^2 (\mu_{nw} - \mu_w) \right]$, and $\mu_{w}$ is the wetting fluid dynamic shear viscosity. We therefore obtain the following expression for the permeability of the tube to the non-wetting fluid:
\begin{equation}
k' = k \left( \frac{\mu_{nw}}{\mu_{w}} \right) \left[2 - \left(2 - \frac{\mu_{w}}{\mu_{nw}} \right) \left(1-\frac{h}{a}\right)^2 \right].
\label{eq:PermNWS}
\end{equation}
Assuming that the permeability of the overall stratum can be described as a bundle of such capillaries in parallel \cite{bear2013dynamics}, and assuming that the stratum permeability is primarily controlled by the pore throat size (with minimal change to the porosity due to the residual fluid), we thus apply the expression given by Eq. \eqref{eq:PermNWS} to describe the permeability of each stratum, $k_{i}'$. Substituting the expressions for $p_{c,i}'$ and $k_{i}'$ in Eq. \eqref{eq:Cacrflow} then yields Ca$^{**}$, the threshold value of Ca for secondary imbibition. We note, however, that for simplicity we have assumed that the thin wetting films are stable and uniform. The real picture, however, can be more complex: these wetting films can become unstable to perturbations, which lead to a Rayleigh--Plateau type instability of the film, and eventually can lead to formation of bridges across the pore \cite{Goren62,Hammond83,Zhao18,Primkulov20}, disrupting the flow and making it a simultaneous two-phase flow \cite{Joseph97}. Investigating how these complexities alter the dynamics of secondary imbibition will be an interesting direction for future work.

\section{Acknowledgements}
%%%%%%%%%%%%%%% ACKNOWLEDGEMENTS
It is a pleasure to acknowledge Gary Hunter, Hubert King, and Jeremy Brandman from ExxonMobil Corporate Strategic Research for stimulating discussions. This work was supported by ExxonMobil through its membership in the Princeton E-filliates Partnership of the Andlinger Center for Energy and the Environment, and in part by the Grand Challenges Initiative of the Princeton Environmental Institute. N.B.L. was also supported in part by the Mary and Randall Hack Graduate Award of the Princeton Environmental Institute. 

\textbf{Author contributions:} N.B.L. and S.S.D. designed the experiments; N.B.L. performed experiments; N.B.L. and S.S.D. analyzed the data; A.A.P. developed the theoretical model through discussions with S.S.D., N.B.L., C.A.B., D.B.A., and H.A.S.; S.S.D. designed and supervised the overall project. All authors discussed the results and wrote the manuscript.


\end{document}